\documentclass[sigplan,screen]{acmart}

\AtBeginDocument{%
  \providecommand\BibTeX{{%
    \normalfont B\kern-0.5em{\scshape i\kern-0.25em b}\kern-0.8em\TeX}}}

\setcopyright{acmcopyright}

\acmYear{2022}\copyrightyear{2022}
\setcopyright{rightsretained}
\acmConference[PPoPP '22]{27th ACM SIGPLAN Symposium on Principles and Practice of Parallel Programming}{February 12--16, 2022}{Seoul, Republic of Korea}
\acmBooktitle{27th ACM SIGPLAN Symposium on Principles and Practice of Parallel Programming (PPoPP '22), February 12--16, 2022, Seoul, Republic of Korea}
\acmPrice{}
\acmDOI{10.1145/3503221.3508425}
\acmISBN{978-1-4503-9204-4/22/02}

\usepackage{tikz}
\usepackage{graphicx}
\usepackage{subcaption}
\usepackage{multirow}
\usepackage{bm}
\usepackage{mhchem}[version=4]
\usepackage{nicefrac}

\acmSubmissionID{134}

\begin{document}

\title{Extending the limit of molecular dynamics with \textit{ab initio} accuracy to 10 billion atoms}




\author{
    Zhuoqiang Guo$^{1,2}$,\,
    Denghui Lu$^4$,\,
    Yujin Yan$^{1,2}$,\,
    Siyu Hu$^{1,2}$,\,
    Rongrong Liu$^{1,2}$,\,
    Guangming Tan$^{1,2,3}$,\,
    Ninghui Sun$^{1,2,3}$,\,
    Wanrun Jiang$^5$,\,
    Lijun Liu$^6$,\,
    Yixiao Chen$^7$,\,
    Linfeng Zhang$^{5,8}$,\,
    Mohan Chen$^4$,\,
    Han Wang$^{9}$,\,
    Weile Jia$^{1,2,3,*}$
}
\affiliation{
 \institution{$^1$Institute of Computing Technology, Chinese Academy of Sciences,\, Beijing,\, China}
 \city{}
 \country{}
 }
\affiliation{
 \institution{$^2$ University of Chinese Academy of Sciences,\, Beijing,\, China}
 \city{}
 \country{}
 }
\affiliation{
 \institution{$^3$ State Key Laboratory of Computer Architecture, ICT, Chinese Academy of Sciences,\, Beijing,\, China}
 \city{}
 \country{}
 }

\affiliation{
 \institution{Email:\,
    \{guozhuoqiang20z, yanyujin17z, husiyu20b, 
    liurongrong21s, tgm, snh, jiaweile\}@ict.ac.cn,\,
    }
 \city{}
 \country{}
 }


\affiliation{
 \institution{$^4$HEDPS, CAPT, College of Engineering, Peking University,\, Beijing,\, China}
 \city{}
 \country{}}
 
\affiliation{
 \institution{%
    Email:\,\{denghuilu,mohanchen\}@pku.edu.cn}
 \city{}
 \country{}}
 
 
\affiliation{
 \institution{$^5$AI for Science Institute, \, Beijing,\, China}
 \city{}
 \country{}}
 
\affiliation{
 \institution{%
    Email:\,wanrunj@live.com}
 \city{}
 \country{}}

\affiliation{
 \institution{$^6$Osaka University,\, Osaka,\, Japan}
 \city{}
 \country{}}

\affiliation{
 \institution{%
    Email:\,liu@mech.osaka-u.ac.jp}
 \city{}
 \country{}}
    
\affiliation{
 \institution{$^7$Princeton University,\, Princeton,\, USA}
 \city{}
 \country{}}
 
\affiliation{
 \institution{%
    Email:\,yixiaoc@princeton.edu}
 \city{}
 \country{}}

\affiliation{
 \institution{$^8$DP Technology, Beijing, China}
 \city{}
 \country{}}

\affiliation{
 \institution{%
    Email:\,zhanglf@dp.tech}
 \city{}
 \country{}}

\affiliation{
 \institution{$^{9}$Laboratory of Computational Physics, Institute of Applied Physics and Computational Mathematics,\, Beijing,\, China}
 \city{}
 \country{}}

\affiliation{
 \institution{%
    Email:\,wang\_han@iapcm.ac.cn}
 \city{}
 \country{}}

\renewcommand{\shortauthors}{Zhuoqiang Guo,Denghui Lu, et al.}

\newcommand{\Or}{\mathcal{O}}
\newcommand{\jump}[1]{\big[\hspace{-0.7mm} \big[ #1 \big]
  \hspace{-0.7mm} \big]}
\newcommand{\mean}[1] {\big\{ \hspace{-0.7mm} \big\{ #1 \big\}
  \hspace{-0.7mm} \big\}}
\newcommand{\abs}[1]{\left\lvert#1\right\rvert}
\newcommand{\norm}[1]{\left\lVert#1\right\rVert}
\newcommand{\average}[1]{\left\langle#1\right\rangle}
\newcommand{\bra}[1]{\langle#1\rvert}
\newcommand{\ket}[1]{\lvert#1\rangle}
\newcommand{\mc}[1]{\mathcal{#1}}
\newcommand{\DG}{\mathrm{DG}}
\newcommand{\ud}{\,\mathrm{d}}

\newcommand{\bd}[1]{\boldsymbol{#1}}
\newcommand{\wt}[1]{\widetilde{#1}}
\newcommand{\wh}[1]{\widehat{#1}}
\newcommand{\wb}[1]{\overline{#1}}

\newcommand{\onlinecite}[1]{\citenum{#1}}
\newcommand{\bvec}[1]{\mathbf{#1}}
\newcommand{\vr}{\bvec{r}}
\newcommand{\vR}{\bvec{R}}
\newcommand{\I}{\mathrm{i}} 

\newcommand{\vace}{\widetilde{V}_{\mathrm{X}}}
\newcommand{\kace}{K^{\mathrm{ACE}}}
\newcommand{\ext}{\mathrm{ext}}
\newcommand{\Hxc}{\mathrm{Hxc}}
\newcommand{\X}{\mathrm{X}}
\newcommand{\xc}{\mathrm{xc}}
\newcommand{\eff}{\mathrm{eff}}

\newcommand{\recheck}[1]{\textcolor{red}{#1}}
\newcommand{\WL}[1]{\textcolor{cyan}{[WL:#1]}}
\newcommand{\LZ}[1]{\textcolor{blue}{[LZ:#1]}}
\newcommand{\WH}[1]{\textcolor{blue}{[WH:#1]}}
\newcommand{\LuDh}[1]{\textcolor{magenta}{#1}}
\newcommand{\MC}[1]{\textcolor{brown}{#1}}
\newcommand{\YJ}[1]{\textcolor{purple}{[YJ:#1]}}
\newcommand{\YC}[1]{\textcolor{green}{[YC:#1]}}
\newcommand{\ZQ}[1]{\textcolor{magenta}{[ZQ:#1]}}

\newcommand{\REV}[1]{{\color{red}#1}}


\begin{abstract}

High-performance computing, together with a neural network model trained from data generated with first-principles methods, has greatly boosted applications of \textit{ab initio} molecular dynamics in terms of spatial and temporal scales on modern supercomputers. Previous state-of-the-art can achieve $1-2$ nanoseconds molecular dynamics simulation per day for 100-million atoms on the entire Summit supercomputer. In this paper, we have significantly reduced the memory footprint and computational time by a comprehensive approach with both algorithmic and system innovations. The neural network model is compressed by model tabulation, kernel fusion, and redundancy removal. Then optimizations such as acceleration of customized kernel, tabulation of activation function, MPI+OpenMP parallelization are implemented on GPU and ARM architectures. Testing results of the copper system show that the optimized code can scale up to the entire machine of both Fugaku and Summit, and the corresponding system size can be extended by a factor of $134$ to an unprecedented $17$ billion atoms. 
The strong scaling of a $13.5$-million atom copper system shows that the time-to-solution can be {7} times faster, reaching {$11.2$} nanoseconds per day. This work opens the door for unprecedentedly large-scale molecular dynamics simulations
based on {\it ab initio} accuracy and can be potentially utilized in studying more 
realistic applications such as mechanical properties of metals, semiconductor devices, 
batteries, etc.
The optimization techniques detailed in this paper also provide
insight for relevant high-performance computing applications.

\end{abstract}





\maketitle
\begingroup\renewcommand\thefootnote{*}
\footnotetext{Corresponding author}
\endgroup

\section{Introduction} \label{sec:introduction}

Machine learning (ML) applications are playing an increasingly important role 
in modern high-performance computing (HPC) systems. 
Besides the optimization of gigantic neural network model training
~\cite{sc2020gems, bian2021arxiv, floridi2020gpt, rajbhandari2020zero, li2021chimera}, 
HPC plus artificial intelligence (AI) in solving scientific problems are 
gaining momentum~\cite{gb2020special,2020Pushing, karimpouli2020physics, jumper2021highly}. 
One example is the machine-learning 
molecular dynamics (MLMD), which aims to bridge the gap between 
first-principles accuracy and Newtonian MD efficiency
~\cite{friederich2021machine, zhang2018deep}. 
In MLMD, the atomic potential energy surface (PES) 
is fitted to obtain \textit{ab initio} accuracy via ML-based models trained 
from first-principles data, which can be generated from density 
functional theory (DFT) calculation software packages such 
as VASP, Quantum Espresso, PWmat, etc
~\cite{hacene2012accelerating, hutchinson2012vasp, JIA2013, JIA2013102, RomeroJoshua2018}. 
Many machine-learning models, such as the Gaussian regression
~\cite{bartok2010gaussian, szlachta2014accuracy}, 
linear regression~\cite{podryabinkin2017active, jinnouchi2019fly,THOMPSON2015316,PhysRevB2019Drautz}, 
and neural network methods~\cite{ behler2007generalized, zhang2018end},
can be applied in the MLMD. 
The resulting MLMD packages, such as the representative works listed in 
Table~\ref{tab:soaSP}, can significantly increase the spatial and 
temporal limit of AIMD. 
One state-of-the-art MLMD code is the open-source package DeePMD-kit~\cite{wang2018kit}, 
which adopts Deep Potential~\cite{zhang2018deep,zhang2018end}, a neural network approach by combing the symmetry-preserving features 
and highly efficient implementations. Compared with the best reported 
AIMD results, the spatial and temporal scales of the accommodated physical
systems can speed up by a factor of 100 and 1000, respectively~\cite{2020Pushing}. 
Recently, the size of the system that the DeePMD method can handle reaches $100$-million atoms on the entire Summit supercomputer (2020 Gordon Bell prize), achieving $91/171/285$ PFLOPS for double/mixed-single/mixed-half precision arithmetic operation. This opens the door for tackling important scientific problems with unprecedented system size and time scales. For example, recent works that used DeePMD-kit focused on the interactions of molecules in water~\cite{galib2021reactive}, nucleation of liquid silicon~\cite{bonati2018silicon}, phase transition of water~\cite{gartner2020signatures}, and phase diagram of water~\cite{zhang2021phase}, etc.

\begin{table*}[]
    \centering
    \caption{Performance of MLMD applications with \textit{ab initio} accuracy. The abbreviations TtS, BP, and DP stand for time-to-solution, Behler-Parrinello scheme, and Deep Potential, respectively. 
    *The parallel efficiency does not significantly decay at the largest machine scale tested in the work, so it is highly likely that they can scale to larger machines.
    $\dagger$Vienna Scientific Cluster (VSC), an HPC system with Intel Xeon Gold 6138 CPUs.
    $\ddagger$An unknown cluster with Intel Xeon E5-2650v2 CPUs at the KISTI supercomputing center.
    **The baseline is the current state-of-the-art DeePMD-kit.}
   \resizebox{\textwidth}{\height}{
   \begin{tabular}{l c c c  r r r r r @{\hspace{3em}}l l}
    \toprule
    Work & Year & Pot. & System & \# atoms & \# CPU {\small cores} & \# GPUs & Machine & \multicolumn{1}{c@{\hspace{2.5em}}}{Peak[{\small FLOPS}]} & \multicolumn{2}{@{\hspace{-2em}}r}{TtS [{\small s/step/atom}]}\\ \midrule
    Simple-NN~\cite{lee2019simple}* & 2019 & BP & \ce{SiO$_2$} & 14K & 80 & -- & Unknown$\ddagger$ & ? & $3.6\times 10^{-5}$  \\
    Singraber {\small el.al.}~\cite{singraber2019library}* & 2019 & BP & \ce{H$_2$O} & 9K & 512 & -- & VSC$\dagger$ & ? & $1.3\times10^{-6}$  \\
    Baseline~\cite{2020Pushing}**(double)     & 2020 & DP & Cu & 127M & 27.3K & 27.3K & Summit & 91P  & $8.1\times 10^{-10}$  \\   
    Baseline~\cite{2020Pushing}**(mixed-half) & 2020 & DP & Cu & 127M & 27.3K & 27.3K & Summit & 275P & $2.7\times 10^{-10}$  \\   
      \midrule
    This work (double) & 2021 & DP & Cu & 3.4B & 27.3K & 27.3K & Summit & 43.7P & $1.1 \times 10^{-10}$  \\   
    This work (double) & 2021 & DP & Cu & {17B} & {7,630K} & - & Fugaku &  {119P}  & $4.1 \times 10^{-11}$  \\
    \bottomrule
    \end{tabular}
    }
    \label{tab:soaSP}
\end{table*}

Despite the success, MLMD software packages still face challenges on modern HPC platforms. First, the diversity of many-core architecture 
supercomputers has raised the issue of performance portability. For example, 
the efficient MPI+CUDA implementation of DeePMD-kit cannot be adopted
to fully exploit the computing power of Fugaku, which uses many-core ARM CPU and currently ranks No. 1 on the top 500 list~\cite{Top500list}. 
It remains unknown whether the optimal granularity and data parallelism on 
GPU works on many-core CPU architecture for HPC+AI applications. 
Second, it is unclear whether the current DeePMD-kit implementation is the optimal choice. For example, model compression techniques, 
such as pruning and low-rank factorization
~\cite{guo2020accelerating, han2015deep, hubara2017quantized, choudhary2020comprehensive}, 
are introduced in the 
fields of natural language processing and computer vision. The compressed model, as discussed in Ref.~\cite{choudhary2020comprehensive}, may suffer from accuracy loss by throwing away some parameters. For scientific computing applications such as the MLMD, 
the difficulty lies in compressing the neural network model without loss of accuracy. 
Third, for problems in complex chemical reactions, electrochemical cells, 
nanocrystalline material, radiation damage, dynamic fracture and crack propagation, 
etc., the required spatial and temporal scales can even go beyond $100$ million atoms and several nanoseconds. Thus, extending the limit of MD with \textit{ab initio} accuracy is important for many scientific applications. 
In this paper, we attempt to solve the above problems by algorithmic and system innovations. It is noted that a flat MPI version of DeePMD-kit~\cite{2020Pushing} (current state-of-the-art) is set as the baseline throughout this paper.

The main contributions of this paper are: 
\begin{itemize}
    \item We introduce a novel algorithm of tabulating the neural 
    network model with fifth-order polynomials, and 
    save $82\%$ of floating-point operations (FLOPs) 
    compared with the previous state-of-the-art. 
    \item We optimize the memory usage and computational time of the 
    embedding matrix, which take more than $95\%$ of total memory footprint, 
    by applying system optimizations such as kernel fusion, redundancy removal, etc. 
    \item We implement an optimized GPU version of DeePMD-kit on Summit.
    Testing results show that our optimized code can be 3.7 to 9.7 
    times faster, and the physical system size can be {$26.7$} times
    bigger ({3.4 billion} atoms) compared to the current state-of-the-art. 
    \item We implement an optimized version of DeePMD-it on Fugaku, 
    and obtained 21.2 and 46.7 times of
    speedup for water and copper system, respectively. 
    Normalized testing results with respect to peak performance and power consumption show that our Fugaku implementation can be ~$1.2$ and ~$1.3$ times faster than
    V100 GPU, respectively. 
    \item Weak scaling shows that the optimized DeePMD-kit can scale up to the 
    entire Summit and Fugaku, reaching an unprecedented {25 and 17} billion atoms
    of water and copper, respectively.
    The strong scaling on Summit shows that DeePMD-kit can reach 
    {11.24} nanoseconds MD simulation per day for a {13.5}-million atom copper system. 
    Such spatial and temporal scales further extend
    the capability of MD with \textit{ab intitio} accuracy.
\end{itemize}

The rest of this paper is organized as follows: The Deep Potential algorithm is 
introduced in Sec.~\ref{sec:algorithm},
with algorithmic and system innovations  provided in Sec.~\ref{sec:innovation}. 
The physical system and testing platform are presented in Sec.~\ref{sec:system}
and \ref{sec:machine}, respectively. Results are discussed and analyzed in 
Sec.~\ref{sec:performance}. 
Conclusions are drawn in Sec.~\ref{sec:conclusion}.  

\section{DEEP POTENTIAL MODEL}\label{sec:algorithm}

The Deep Potential (DP) model is constructed with deep neural networks for representing the potential energy surface in a symmetry preserving manner. 
It achieves comparable accuracy as the~\textit{ab initio} calculations and reduces the computational complexity to linearly depending on the degrees of freedom in the system. 
In this section, we introduce the construction of the DP model and analyzing the most expensive part in terms of floating-point operation and memory usage.


\subsection{Model definition}\label{sec:model_define}
\begin{figure*}
    \centering
    \includegraphics[width=0.85\textwidth]{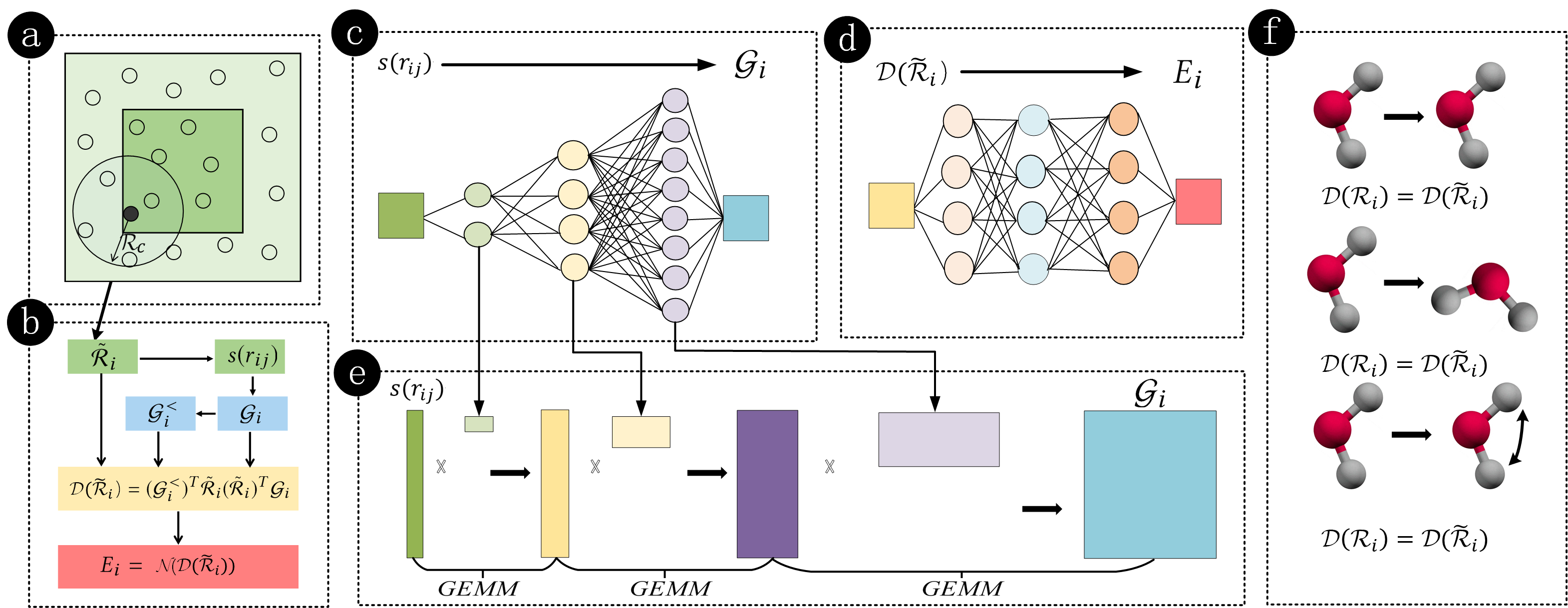}
    \caption{A schematic illustration of the DP model. (a). One sub-region on a single MPI task (Green: local region, light cyan: ghost region). (b). The mapping from local environment $\tilde{\mathcal{R}_i}$ to the energy $E_i$ for a single atom. (c). Embedding net with three hidden layers. (d). Fitting net with three hidden layers. (e). Execution of the embedding net of the previous DP model via matrix-matrix multiplications, the first layer expands vector $s(r_{ij})$ by a factor of $d_1$, the next two layers expand the input matrix by a factor of two. (f).  Physical symmetry preserving descriptors. 
    }
    \label{fig:model}
\end{figure*}

A schematic illustration of the DP model is shown in Fig.~\ref{fig:model}. In a physical system of $N$ atoms, each atom is described by using its atomic position $\bm{r_i}=(x_i,y_i,z_i)\in\mathbb R^3$, $i = 1,2,\dots,N$. 
Each MPI task holds a sub-region of the physical system. 
Note that the DP model assumes the potential energy $E_i$ of atom $i$ only depends on its neighbors $\mathcal{R}_i=\{\bm{r_{ij}}|j\in L_{R_c}(i)\}$, where $\bm{r_{ij}}=\bm{r_j}-\bm{r_i}$ and $L_{R_c}(i)$ denotes the index set of the neighboring atoms within the cut-off radius $R_c$ as shown in  Fig.~\ref{fig:model}~(a). 
In the execution of the DP model, first, the \textit{environment matrix} $\Tilde{\mathcal{R}}_i$ is generated from the neighboring list of a particular atom $i$. 
Then $s(\bm{r_{ij}})$, the first column of the environment matrix, is passed to a three hidden layer \textit{embedding net} to obtain the \textit{embedding matrix} $\mathcal{G}_i$. 
Next, the symmetry-preserving descriptor $\mathcal{D}({\Tilde{\mathcal{R}}_i})$ is constructed, followed by a three-hidden-layer \textit{fitting net} to produce $E_i$, the potential energy of atom $i$.
In the end, the total energy of the system $E$ is calculated by summing each {individual} potential energy, i.e., $E = \sum_i E_i$. 
Note that the $E_i$ is evaluated {in forward propagation}, while the atomic force $F_i$, which is the gradient of the potential energy, is calculated {in the backward propagation}.

A detailed mathematical view of the model is the following:
The input of the DP model is the environment matrix of each atom $i$, which is denoted as $\tilde{\mathcal{R}_i}\in\mathbb{R}^{N_m\times4}$, where $N_m$ is the maximum of all neighbor list.
$\tilde{\mathcal{R}_i}$ is constructed from the local environment, i.e.~all relative postitions of neighboring atoms $\bm{r_{ij}}=\bm{r_j}-\bm{r_i}=\{x_{ij}, y_{ij}, z_{ij}\}$, with the following equations: 
\begin{equation}
    \tilde{\mathcal{R}_i}=s(r_{ij})\times(1,x_{ij}/{\vert \bm{r_{ij}}\vert}, y_{ij}/{\vert \bm{r_{ij}}\vert}, z_{ij}/{\vert \bm{r_{ij}}\vert}),
\end{equation}
where $s(r_{ij})=\omega(\vert \bm{r_{ij}} \vert)/{\vert \bm{r_{ij}}\vert}$ and $\omega(\vert \bm{r_{ij}} \vert)$ is a gating function that decays smoothly from $1$ to $0$ when $\vert \bm{r_{ij}}\vert \leq R_c$.
The DP model is unique in automatically generated descriptors $\mathcal{D}({\Tilde{\mathcal{R}}_i})$, which preserved physical symmetries such as translational, rotational, and permutational invariance:
\begin{equation}
 \mathcal{D}(\Tilde{\mathcal{R}}_i)=(\mathcal{G}_i^<)^T\tilde{\mathcal{R}_i}(\tilde{\mathcal{R}_i})^T\mathcal{G}_i,   
\end{equation}
where $\mathcal{G}_i \in \mathbb{R}^{N_m\times M}$ is the embedding matrix as illustrated in Fig.~\ref{fig:model}(b).
The matrix $\mathcal{G}_i$ obtained by forwarding vector $s(r_{ij})$ through the $m$-layer fully connected embedding net (see Fig.~\ref{fig:model}~(c)):
\begin{equation}
    \mathcal{G}_i=\mathcal{L}_m^e \circ \dots \circ \mathcal{L}_1^e \circ \mathcal{L}_0^e(s(r_{ij})).
\end{equation}
The first layer is a standard fully-connected layer with activation function $\tanh$:
\begin{equation}
   \mathcal{L}_0^e(x)=\tanh(x \cdot W_0^e+b_0^e),\quad W_0^e \in \mathbb R^{d_1},b_0^e \in \mathbb R^{d_1}.
   \label{act_fun1}
\end{equation}
The input size is expanded by $d_1$ times after the first layer. The rest layers are fully-connected layers with shortcut connection and $\tanh$ activation function:
\begin{equation}
    \mathcal{L}_k^e(x)=(x,x)+\tanh(x \cdot W_k^e+b_k^e).
    \label{act_fun2}
\end{equation}
The output size of each layer is doubled and final output size is $M$ (see Fig.~\ref{fig:model}~(e), note $M=4d_1$). The matrix $\mathcal{G}_i^< \in \mathbb{R}^{N_m \times M^<}$ with $M^< < M$ is a sub-matrix of $\mathcal{G}_i$ formed by taking the first $M^<$ columns of $\mathcal{G}_i$. 

The symmetry-preserving descriptor $\mathcal{D}$ is mapped to ato-mic energy $E_i$ via a fitting network $\mathcal{N}$: $E_i=\mathcal{N}(\mathcal{D}(\Tilde{{\mathcal R}}_i))$ (see Fig.~\ref{fig:model}~(d)). 
The fitting network $\mathcal N$ is a standard fully-connected network with hidden layers being of the same size.
A shortcut connection is established between the input and output of each hidden layer of the fitting network. 
The force on atom $i$ is derived from the negative gradient of the total energy: $F_i=-\nabla_{r_i}E=-\sum_j\nabla_{r_i}E_j$.

\subsection{Computationally intensive parts and memory footprint}\label{sec:computational_intensive}

Both the training and inference of the DP model are implemented\, in\, an\, open-source\, package\, DeePMD-kit\cite{zhang2018deep}.\,
{DeePMD-kit is further interfaced with the LAMMPS package~\cite{plimpton1995fast} to perform large-scale MD simulations.}
The model training generally takes a few hours to one week on a single GPU, depending on the complexity of the physical system~\cite{zhang2018end}. 
The model inference, on the other hand, can take hours, even days on supercomputers for simulating a large system with long time scales. 
Thus, in this work, we focus on the computationally intensive part, i.e., the inference of the DP model.

TensorFlow is selected as the framework for building the DP model.  All procedures listed in Fig.~\ref{fig:model} are implemented either with standard TensorFlow operators or with customized TensorFlow operators. For example, both embedding net and fitting net are constructed with standard TensorFlow operators, and the environment matrix $\tilde{\mathcal{R}_i}$ is built via hand-crafted TensorFlow operators.

The most computationally intensive part of the DP model lies in attaining the embedding matrix from $s(r_{ij})$ via the embedding net, as illustrated in Fig.~\ref{fig:model}~(e). Profiling of the previous DeePMD-kit shows that more than $90$ percent of the total time are spent on execution of the embedding net. The input vector $s(r_{ij})$, which is a smooth function of $\frac{1}{|r_{ij}|}$(see Sec.~\ref{sec:model_define}), is expanded from size $N_m$ to $N_m\times 4d_1$
by matrix-matrix multiplication operations, where $N_m$ is the max length of all atomic neighbor lists and $d_1$ is the width of the first fully connected layer.
Theoretically, the total FLOPs of the embedding net is $N_{a}\times (N_m\times d_1 + 10\times N_m \times d_1^2)$, where $N_a$ stands for the number of atoms residing on single MPI task. 
For example, $N_a$ can be up to $4,600$ and $N_m$ is $512$ for the copper system in the previous work\cite{2020Pushing}, and the corresponding width of the three fully connected layers in the embedding net are 32, 64, and 128, respectively. The computational cost of the embedding net approximately accounts for {$95\%$} of the total FLOPs. 
It is noted that the embedding matrix $\mathcal{G}_i$ (dimension: $N_a$$\times$$N_m$$\times$$128$) is the most memory-demanding variable in the DeePMD-kit. For example, $\mathcal{G}_i$ can take $2.4$GB GPU global memory for a system of $4,600$ copper atoms mentioned above. Since several copies of $\mathcal{G}_i$ are required during the model inference, such as in forward and backward propagation, trading time with space, etc., the $\mathcal{G}_i$ related memory consumption accounts for more than $95\%$ of the total. 

\section{INNOVATION} \label {sec:innovation}

\subsection{Summary of contributions}

The main contribution of this paper is a performance portable DeePMD-kit code, which 
further extends the system size and time scale of molecular dynamics with
\textit{ab initio} accuracy on modern HPC platforms. 
The optimization strategy and implementation details are summarized on both 
Fugaku and Summit, providing insights to the optimization of other HPC applications. 



\subsection{Algorithmic innovation} \label{sec:algorithm_innovation}

As discussed in Sec.~\ref{sec:computational_intensive}, the propagation of the embedding net accounts for {$95\%$} of the FLOPs in the baseline implementation, and takes {more than $90\%$ } of the total time. 
The reason, as shown in Fig.~\ref{fig:model}~(e), lies in the computational expense of matrix-matrix multiplications, which rapidly increases through each layer of the embedding net. 
The embedding matrix $\mathcal{G}_i$ is expanded from one vector $s(r_{ij})\in \mathbb R^{N_m}$ by $M$ times in this process, and reaches $N_m\times M$ after the propagation of the embedding net.
Essentially, embedding net can be viewed as a mapping function 
$g:\mathbb{R} \to \mathbb{R}^{M}$, 
from each component of the vector $s({r_{ij}})$ to one row of the embedding matrix $\mathcal{G}_i$. 
By employing Equations \ref{act_fun1} and \ref{act_fun2} in Sec.~\ref{sec:model_define} 
as activation functions, the embedding net is a high-dimensional and extraordinarily complex continuous function\cite{lu2020mean}.

\newtheorem{thm}{\bf Weierstrass Approximation Theorem}
\begin{thm} \label{thm}
Suppose $g$ is a continuous real-valued function defined on the real interval $[a, b]$. For every $\varepsilon > 0$, there exists a polynomial $f$ such that for all $x$ in $[a,b]$, we have $\vert f(x)-g(x) \vert < \varepsilon$, or equivalently, the supremum norm $\|f(x)-g(x) \| < \varepsilon$.
\end{thm}

Based on the Theorem \ref{thm}, polynomials can be introduced to approximate the embedding network.
The domain of the input is equally divided into $n$ intervals with 
nodes denoted by $x_0<x_1<\dots<x_n$. 
In the $\theta$-th interval $(x_{\theta-1}, x_\theta], \theta = 1,2,\dots,n$, we approximate the embedding net with $M$ fifth-order polynomials
$$
f_\theta^\eta(x)=a_{\theta0}^\eta+a_{\theta1}^\eta x+a_{\theta2}^\eta x^2+a_{\theta3}^\eta x^3+a_{\theta4}^\eta x^4+a_{\theta5}^\eta x^5.
$$
In order to determine the coefficients $a_{\theta \xi}^\eta$, $\xi=0,1,2,3,4,5$, $\eta=1,2,\dots,M$, 
we acquire the values, first and second order derivatives of $f_\theta$ and $g$ matched at the nodes $x_{\theta-1}$ and $x_\theta$.
The obtained coefficients $a_{\theta \xi}^\eta$ are gathered and stored as a table, so that
the evaluation of the original embedding net now can be approximated by the polynomials. 
This procedure is also referred as \emph{tabulation} in this work. 
The accuracy of the tabulation relies on the size of the interval $(x_{\theta-1}, x_\theta]$.
With a decreasing interval size, the error introduced by the approximation would vanish.

\begin{figure}[]
    \centering
     \includegraphics[width=0.5\textwidth]{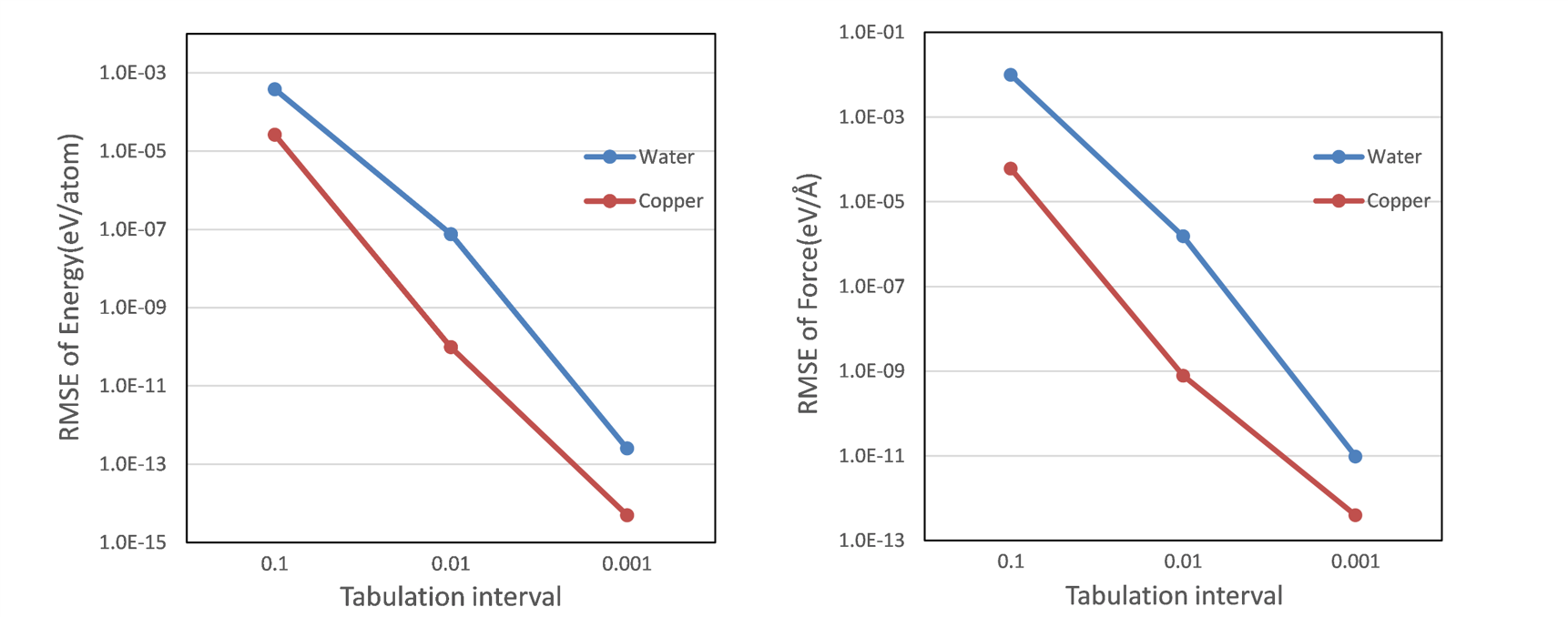}
    \caption{ {Accuracy of the tabulated model compared to the baseline DP model. Root mean squared error (RMSE) of per-atom energy and per-component force are shown, with respect to 3 interval sizes (0.1, 0.01, and 0.001). 100  data are tested for each system}.}
    \label{fig:stride_accuracy}
\end{figure}

Two physical systems, water and copper, are tested to show the accuracy of the tabulated DP model against the baseline in Fig.~\ref{fig:stride_accuracy}. 
The accuracy is measured by the root mean square error (RMSE) in predicted per-atom energy ($RMSE_E$) and per-component force ($RMSE_F$), i.e.,
$$
RMSE_E = \frac{1}{N}{\sqrt{\frac{1}{m}\sum\nolimits_{i=1}^m{(E_i^{tab}-E_i^{orig})^2}},}
$$
and
$$
RMSE_F = {\sqrt{\frac{1}{3mN}\sum\nolimits_{i=1}^{m}\sum\nolimits_{j=1}^{N}\sum\nolimits_{k=1}^{3}(F_{i,j,k}^{tab}-F_{i,j,k}^{orig})^2}},
$$
where $m$ is the number of data to be tested (the number of atomic configurations), $N$ is the number of atoms in each configuration and $k$ represents three directions of Cartesian coordinates.
Quantities with superscripts $tab$ and $orig$ denote the predictions of the tabulated DP model and the original DP model, respectively. 
As shown in Fig.~\ref{fig:stride_accuracy}, the RMSE of the per-atom energy drop from 
{$2.0\times 10^{-5}$ eV/atom to $5.0\times 10^{-15}$} eV/atom and the corresponding per-component force drop from 
{$6.0\times 10^{-5}$ eV/$\Angstrom$ to $4.0\times 10^{-13}$} eV/$\Angstrom$ when the interval varies from $0.1$ to $0.001$. 
Note that the double-precision limit is reached when the interval is set to $0.001$.
This means our tabulated model can be equally accurate as the original model.
We remark that the size of the tabulated model grows as the interval decreases.
For example, the size of the model can be 
{257 MB for water system} when interval equals $0.001$,
but only {33} MB when the interval is $0.01$. Thus in practice, an optimal 
interval is balanced between accuracy and model size, and we choose
$0.01$ as default in our optimized code.
A theoretical analysis of the FLOPs on embedding matrix for both the original and 
tabulated model is as follows. 
We denote the number of atoms by $N_a$, the maximal neighbor list size by $N_m$ and the size of the first layer of the embedding net by $d_1$.
For the purpose of simplicity, the FLOPs of activation function are counted. 
The total number of FLOPs for computing the three-layer embedding net by the original model is
$N_{a}\times (N_m\times d_1+ 10\times N_m \times d_1^2)$, as discussed 
in Sec.~\ref{sec:computational_intensive}, while that for the tabulated model is $N_a\times 56 \times N_m \times d_1$. 
Thus, the theoretical speedup in terms of FLOPs is: $\nicefrac{(1+10\times d_1)}{56}$.
In the previous work\cite{2020Pushing}, $d_1$ is set to be $32$ for both water and copper systems, 
thus the tabulated model saves {$82$} percent of the embedding matrix FLOPs compared to the original model.


\subsection{Optimization strategy} \label{sec:opt_stratergy}

In this section, we will discuss the general strategies used in
optimizing DeePMD-kit on modern many-core architectures. The goal is to 
reduce the memory footprint after the total FLOPs are brought down by
82 percent by the tabulation of embedding net as discussed in
Sec.~\ref{sec:algorithm_innovation}. 

\begin{figure}
    \centering
    \includegraphics[width=0.4\textwidth]{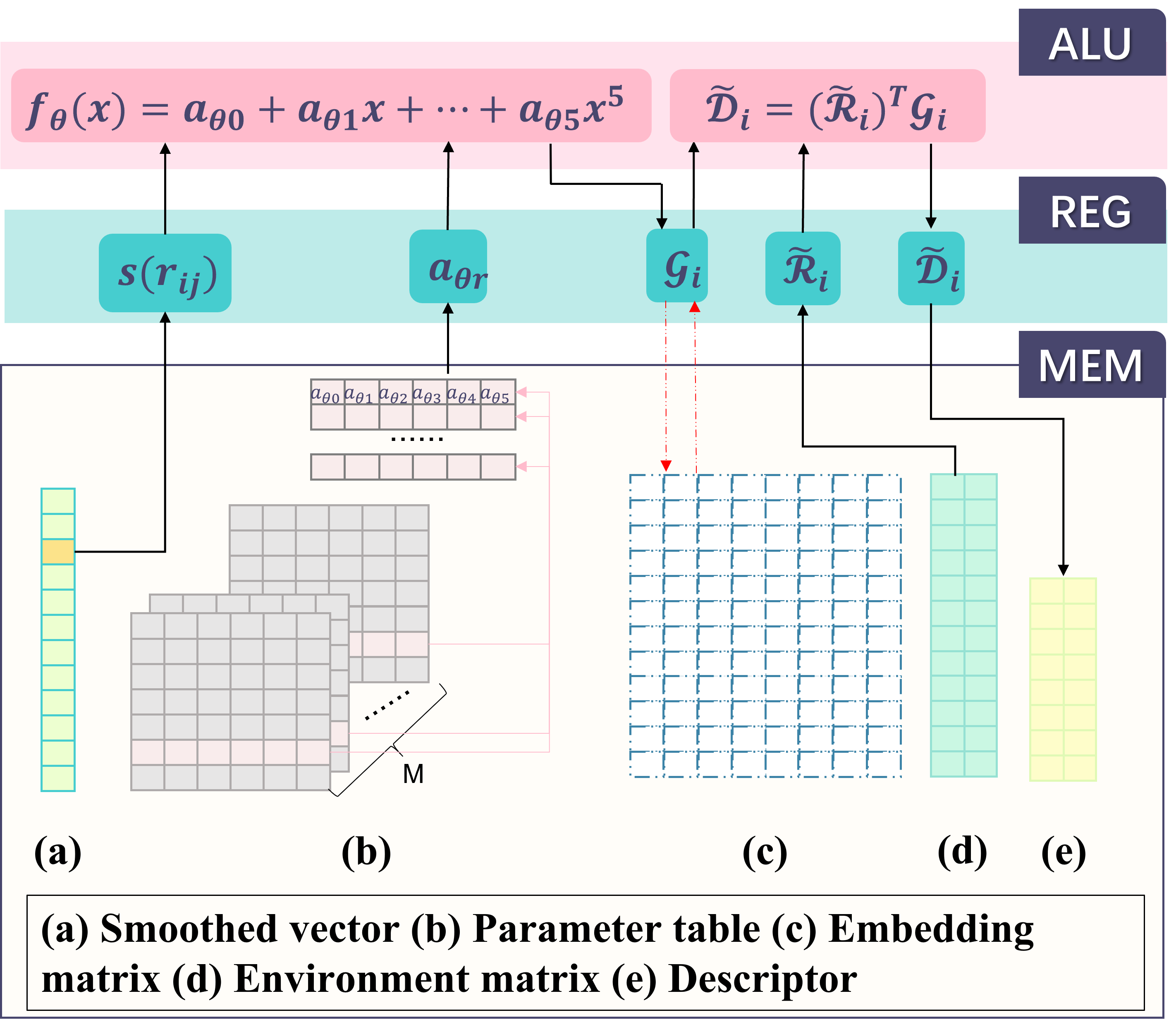}
    \caption{An illustration of the data movement when evaluating the descriptor $\mathcal{D}_i$ of the optimized (tabulated and kernel fused) DP model. Dashed line: memory allocation/access saved by the kernel fusion.
    }
    \label{fig:kernel fusion}
\end{figure}

The first strategy is to reduce the memory footprint by contracting variables and merging
calculations. We find that the embedding matrix $\mathcal{G}_i$, 
as discussed in Sec.~\ref{sec:computational_intensive}, is the most memory-demanding 
variable and takes more than {$95$} percent of the total memory usage in the 
baseline implementation. After the tabulation of the embedding net (See
Sec.~\ref{sec:algorithm_innovation}), $\mathcal{G}_i$ is no longer built through
the three-layer embedding net (Fig.~\ref{fig:model}~(e)), rather, it is approximated 
by fifth-order interpolations. The data movement of evaluating the descriptor 
$\mathcal{D}_i$ on typical modern many-core architecture is illustrated in 
Fig.~\ref{fig:kernel fusion}. For simplicity, we only demonstrate the data movement among
the basic components such as arithmetic logic unit (ALU), register (REG), 
and high-bandwidth memory (MEM). CACHE and shared memory are not discussed. 
The tabulation of embedding net ($f_\theta(x)=a_{\theta0}+a_{\theta1} x
...a_{\theta5} x^5$) loads the vector $s(r_{ij})$ 
and the parameter table, and stores the
embedding matrix  $\mathcal{G}_i$. 
Then $\tilde{\mathcal{R}_i}^T\mathcal{G}_i$ is carried out by calling the matrix-matrix 
multiplication (GEMM) subroutine.  We optimize the  allocation and access of $\mathcal{G}_i$ 
by contracting the tabulation and matrix-matrix multiplication, and the
resulting equation is $\tilde{\mathcal{R}_i}^T f_\theta(s(r_{ij}))$. 
Although the contracted equation saves the load and store of the embedding matrix
$\mathcal{G}_i$, implementation detail still vary depending on the architecture, 
as will be discussed in Sec.~\ref{sec:system_innovation1} and 
Sec.~\ref{sec:system_innovation2}.

\begin{figure}
    \centering
    \includegraphics[width=0.45\textwidth]{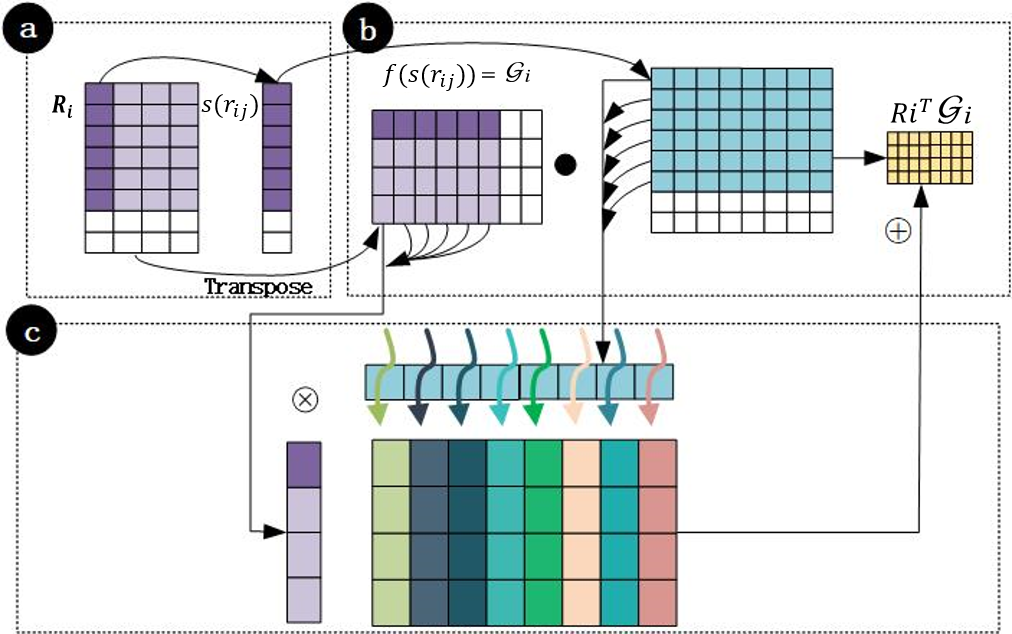}
    \caption{Removing the redundant zeros from the DP model. (a) Redundant zeros in the environment matrix $\mathcal{R}_i$, $s(r_{ij})$. (b) Evaluation of ${\mathcal{\tilde{R}}_i}^T\mathcal{G}_i$.  Note that redundant-zero ratio of the embedding matrix $\mathcal{G}_i$ are related to the $s(r_{ij})$.
    (c) Bypassing redundant zeros in the execution of outer product in one thread block.}
    \label{fig:Redundancy removal}
\end{figure}

Removing redundancy is the second strategy used in our optimization. 
Related work demonstrated that up to 52 percent of performance boost can be gained 
in removing calculations related to redundant zeros~\cite{you2020sc}. 
In the previous DP model, the data redundancy stems 
from the padding of the number of rows of the environment matrix $\mathcal{\tilde{R}}_i$ 
to $N_m$, where $N_m$ is the maximally possible length of atomic neighboring list 
as discussed in Sec.~\ref{sec:algorithm}. Then the padded matrix can take advantage
of the high-performance matrix-matrix multiplication by calling the GEMM subroutine. 
However, since the tabulation of the embedding matrix and the consecutive GEMM
operations are merged as discussed above, the redundancy related operations 
can also be bypassed. Note that data redundancy exists not only in $\mathcal{R}_i$, 
but also in $s(r_{ij})$ and the corresponding embedding matrix 
$\mathcal{G}_i$ (Fig.~\ref{fig:Redundancy removal}~(a)).
The size of $N_m$ can be as large as several hundred, depending on the cutoff radius $R_c$.  
For example, $N_m$ is set to be $128$ and $512$ for the water and copper systems,
respectively~\cite{2020Pushing}. 
The copper model is trained for both ambient and high-pressure conditions, 
while the water model is trained only for ambient conditions. 
The $N_m$ reserved for the copper model is larger because the neighbor lists are 
longer in the high-pressure conditions (higher density and thus more neighbors 
in the cutoff radius). When the models are used under ambient conditions, the copper 
model has a higher degree of redundancy, and the redundancy can not be avoided nor 
reduced due to the data layout and matrix operations in the baseline implementation.

The third optimization strategy is to improve the parallel efficiency by
increasing the computational {granularity} of each individual MPI process. 
We notice that given a fixed physical problem, the ratio between computation 
over communication decreases as we use more MPI tasks. 
This is mainly because the communication volume of the 
ghost region increases when using more MPI tasks. 
For example, in a one-dimensional problem, suppose the size of the ghost region of an 
individual MPI is $V$ and the total number of MPI process is $n$, then the 
communication volume of the ghost region is $n\times V$. This means 
theoretically, it is optimal to launch only one MPI per node and distribute
the workload into finer granularity by using OpenMP, CUDA, etc...
In practice,  MPI+X  is the most frequently used parallelization 
schemes,  and X can be OpenMP, CUDA, HIP, etc.

\subsection{Implementation I: GPU} \label{sec:system_innovation1}

In the following subsections, we will focus on optimizing the DeePMD-kit
on the GPU heterogeneous architecture
based on the strategies outlined in Sec.~\ref{sec:opt_stratergy} . 

\subsubsection{Kernel fusion} \label{sec:kernel_fusion} 

We optimize the GPU memory footprint by merging the tabulation of the embedding net and consecutive matrix-matrix multiplication into a single CUDA customized kernel.
In the optimized DeePMD-kit, each row of $\mathcal{G}_i$ is evaluated 
in one thread block and stored in the register (without storing back to
global memory), then one column of the environment matrix 
$\tilde{\mathcal{R}_i}^T$ is loaded into the registers to perform
an outer product. An illustration of the outer product is shown in Fig.~\ref{fig:Redundancy removal}. 
The result from all thread blocks is added up to from $\tilde{\mathcal{R}_i}^T\mathcal{G}_i$.
Note that the dimension of the outer product matrix is $4\times M$, and can be 
accommodated in the shared memory on V100 GPUs for an efficient summation.  
We remark that $\mathcal{G}_i$ is neither allocated nor moved between global 
memory and registers in  the optimized code, and both the memory footprint 
and computational time are significantly reduced after the kernel fusion, 
as will be discussed in Sec.~\ref{sec:performance}.

\subsubsection{Redundancy removal} \label{sec:redundancy}


The fused CUDA kernel introduced in Sec.~\ref{sec:kernel_fusion} makes 
it possible to bypass the redundant zeros in the DP model.
As shown in Fig.~\ref{fig:Redundancy removal}, one column of 
$\tilde{\mathcal{R}_i}^T$, together with one row of $\mathcal{G}_i$, 
are held by one thread block to perform an outer product. 
The dimension of the resulting matrix 
$\tilde{{\mathcal{D}_i}} = \tilde{\mathcal{R}_i}^T\mathcal{G}_i$ is $4 \times M$, 
where $M$ is typically $128$. Note that if element $j$ in $s(r_{ij})$ is padded, 
the corresponding column $j$ of $\tilde{\mathcal{R}_i}^T$ and 
row $j$ of $\mathcal{G}_i$ are also padded with constant numbers. 
Thus, the corresponding final results can be directly derived without 
any floating-point operations.

\subsubsection{Other optimizations}

The customized TensorFlow operators are also accelerated. For example, 
ProdEnvMatA, which evaluates the environment matrix from the neighbor list of atom $i$, is further optimized on GPU by carefully using shared memory and removing redundancy. 
Testing results show that the optimized version can be 3 times faster compared to the original one. 
A random global memory writing policy is applied to ProdForceSeA and ProdVironSeA, which are operators for computing the forces and virial tensors, and improves the performance of atomic addition.


\subsection{Implementation II: ARM CPU}  \label{sec:system_innovation2}

Starting from a flat MPI version of the DeePMD-kit~\cite{2020Pushing}, we apply 
the optimization strategies listed in Sec.~\ref{sec:opt_stratergy} on the Fugaku 
supercomputer in the following subsections. 


\subsubsection{Tabulation of embedding net} \label{sec:tabulation_layout}
In our previous implementation, the fifth-order tabulated model introduced in Sec.
~\ref{sec:algorithm_innovation} are stored as an array of structures (AoS).
The 6 coefficients of the polynomials are stored continuously as a row. However, AoS
can not fully exploit the {1024 GB/s} bandwidth provided by 
the A64FX CPU due to the discontinuous memory access. We optimize 
the data layout of the tabulated model by transposing every 16 structures 
so that the 512-bit scalable vector extension (SVE) instructions are 
invoked in accessing the table. Note that 16 is chosen because the 
loop is unrolled two times to take advantage of the two floating-point
operation pipeline on A64FX($16\times64=1024$ bits). 
We remark that the data layout of the tabulated model is not 
transposed during the MD simulation, but during the tabulation as part 
of the post-processing.



\subsubsection{Kernel fusion and redundancy removal}\label{sec:kernel_fusion_and_redundancy_removal}

The tabulation of the embedding net and consecutive matrix-matrix multiplication 
are merged into a single function, and then the FLOPs with respect to redundant elements
are avoided on A64FX. The outer product of the environment matrix  $\tilde{\mathcal{R}_i}^T$
and the embedding matrix $\mathcal{G}_i$ is implemented on the ARM CPU to enhance the 64 KB L1 CACHE hit rate. We remark that the implementation of the kernel fusion and redundancy removal follows the same pattern on both A64FX and V100 GPU. 

\begin{figure}[h]
    \centering
    \includegraphics[width=0.45\textwidth]{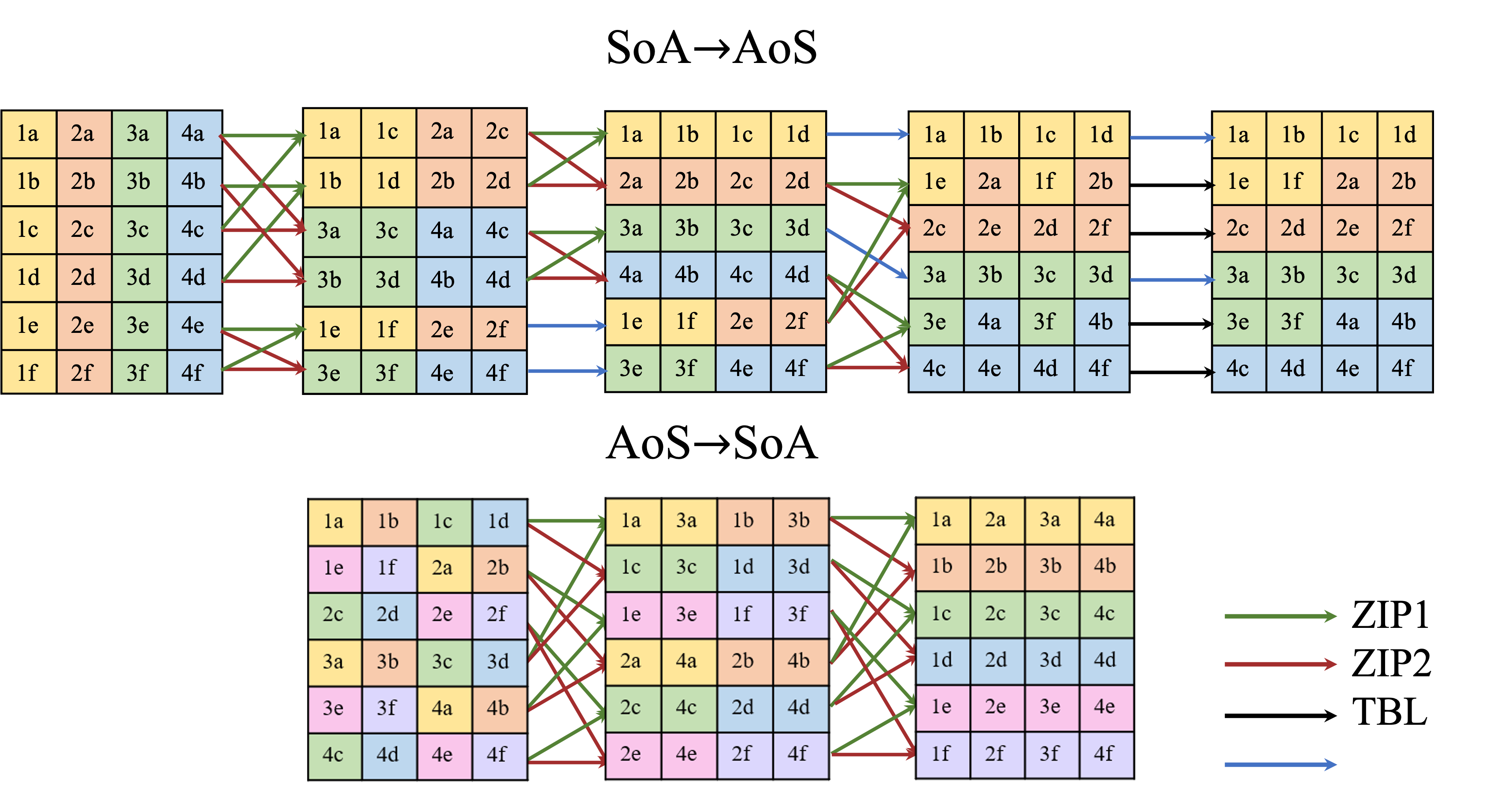}
    \caption{An illustration of converting between AoS and SoA by utilizing SVE instructions. Conversion of 6x4 arrays is plotted for simplicity. 
    Each row is a vector register or is stored continuously in memory.}
    \label{fig:SoA AoS transfer}
\end{figure}

\subsubsection{Other optimizations}\label{sec:vectorization}

We further optimize the customized TensorFlow operators by exploiting vectorization.
One key data structure is descrpt\_a\_deriv 
which is used in multiple customized TensorFlow 
operators such as ProdEnvMatA, ProdViralSeA and ProdForceSeA.  
The variable descrpt\_a\_deriv 
is stored as an AoS, and is required to be converted between SoA and AoS for 
vectorization purposes.  
For AoS of the size of {2$\times$8, 3$\times$8, and 4$\times$8}, the conversion can be performed with single ld2, ld3, or ld4 SVE
instructions, respectively. However, in the DeePMD-kit, the size of descrpt\_a\_deriv is {12$\times$8} and 
can not be trivially converted with single SVE instruction.  
{We implement a fast converting subroutine to switch between SoA and AoS by utilizing 512-bit SVE 
instructions, as illustrated in Fig.~\ref{fig:SoA AoS transfer}. 
Then the corresponding customized TensorFlow operators are unrolled and vectorized to exploit the computing power of A64FX. }

The $\tanh$ function is selected as the activation function for both embedding and fitting net due to 
accuracy considerations~\cite{zhang2018deep}, and it is optimized with a second-order polynomial approximation. 
Only the positive part of the function is tabulated because it is an odd function
(\textit{$\tanh(-x)=-\tanh(x)$}). The upper bound  of $x$ is chosen to be $8$, and
for any $x$ greater than $8$, $\tanh(x)$ is set to $1.0$. Our tabulation of 
the $\tanh$ function can be 60 times faster compared with the original one on A64FX, and the corresponding error is about $1.0\times10^{-7}$. It is noted that our tabulation of the $\tanh$ does not affect the overall accuracy of 
the code. 

\subsubsection{MPI+OpenMP} \label{sec:MPI+openMP}

\begin{figure}
    \centering
    \includegraphics[width=0.45 \textwidth]{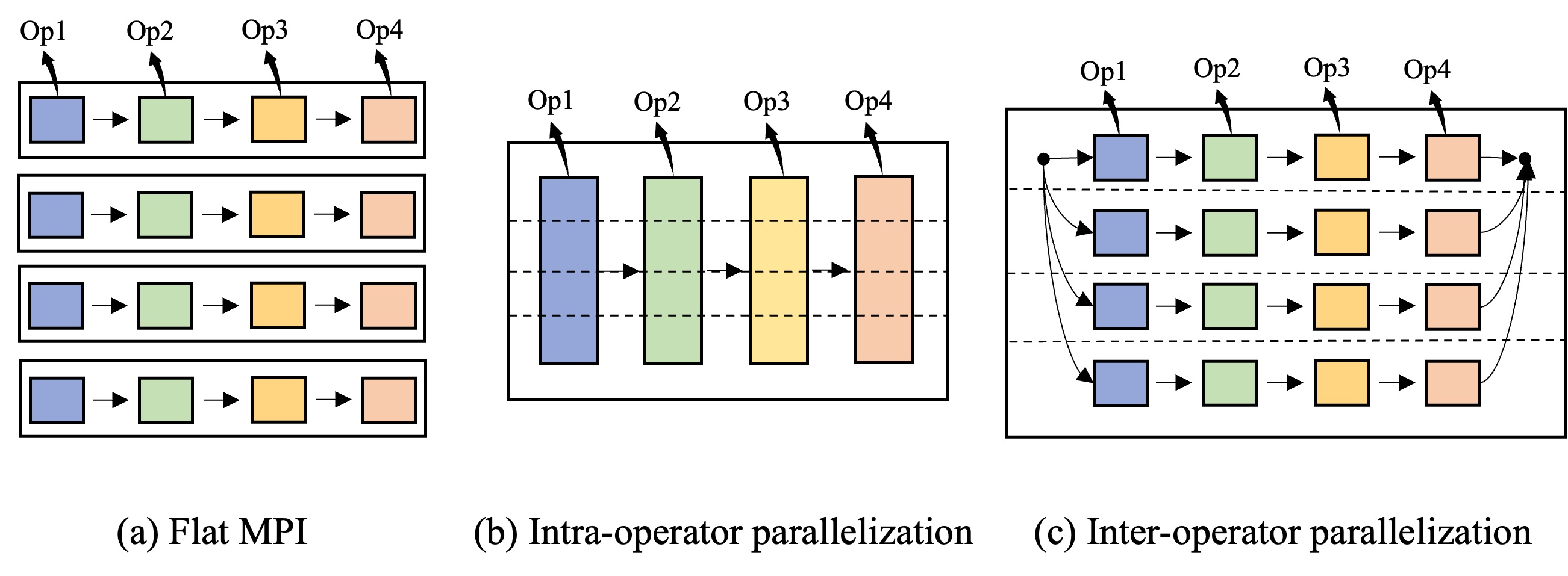}
    \caption{Hybrid paralllelization scheme including (a) Flat MPI parallelization, (b) each TensorFlow operator is parallelized among multiple threads in single MPI, and (c) each OpenMP thread holds a fraction of the sub-region and executed in parallel. }
    \label{fig:MPI+OpenMP}
\end{figure}

Previously, the DeePMD-kit is parallelized on Fugaku in a flat MPI scheme to exploit 
the performance of many-core architecture, as shown in Fig.~\ref{fig:MPI+OpenMP}(a).
However, due to the limited size of high bandwidth memory on A64FX (32GB),
each MPI task can only allocate a maximum of 0.67GB memory (32GB/48=0.67GB). Thus
the size of the sub-region resides on a single MPI task is highly restricted by the 
memory available. Furthermore, the TensorFlow graph, along with MPI buffers, 
are allocated and kept 48 copies on a single Fugaku node and deteriorate 
the situation. A hybrid parallelization scheme using MPI+OpenMP can 
effectively reduce the number of MPI used, increasing parallel granularity 
and reducing inter-MPI communication, as discussed in Sec.~\ref{sec:opt_stratergy}. 
We find that intra-operators MPI+OpenMP parallelization 
(See Fig.~\ref{fig:MPI+OpenMP}(b)) is not efficient due to
frequently forking and joining between different operators. And some customized 
TensorFlow operators, such as ProdVirialSeA, ProdForceSeA,  having writing conflicts 
and can not be easily parallelized with OpenMP. 

Then we implement an inter-operator parallelization scheme, 
as shown in Fig.~\ref{fig:MPI+OpenMP}(c). 
In this scheme, each OpenMP thread mimics
the behavior of an MPI task by holding a fraction of the sub-region. 
Note that the sub-region is carefully divided to avoid load-balance problems, 
and thread forking and joining only occur once per MD step. In each MPI task, 
only one copy of the TensorFlow graph is kept and shared among OpenMP threads, 
and the inter-MPI communication is significantly reduced due to the increase 
of granularity. The optimized code can accommodate systems {1.5} time bigger 
compared to the flat MPI implementation. We remark that the  
performance of the hybrid parallelization scheme is highly
related to the memory affinity of the non-uniform memory access (NUMA) node. 

\section{The physical systems} \label {sec:system}

The performance of the optimized DeePMD-kit is measured on two typical systems: water and copper, which have been extensively trained and studied in 
Refs.~\cite{zhang2018deep,zhang2018end}, and \cite{zhang2020dp}, respectively. 
The cutoff radius of water and copper systems are chosen to be 6 and 8~\AA, 
and the corresponding maximal number of neighbors are 138 and 500, respectively. 
In the previous DP model, the sizes of the embedding net and fitting net are set to be
 $32\times 64 \times 128$ and $240\times240\times240$, respectively. 
 In the optimized DeePMD-kit code, these models are compressed and tested on both Summit
 and Fugaku supercomputers. 

The strong scaling of the optimized DeePMD-kit is tested using the water
system composed of $41,472,000$ and $8,294,400$ atoms on 4,560 computing 
nodes of Summit and Fugaku, respectively, and the weak scaling is 
performed using a copper system with {122,779} and 6,804 atoms per 
MPI task. We remark that our tests do not reach the entire Fugaku
supercomputer due to the computing resource accessible to us. 
The configuration of the water system is made by replicating a well 
equilibrated liquid water system of 192 atoms, and that of the copper system 
are generated as perfect face-centered-cubic (FCC) lattice with the 
lattice constant of 3.634~\AA. 
A total of 99 MD steps evaluated by the Velocity-Verlet algorithm is performed
(the energy and forces are evaluated 100 times), and time steps for water 
and copper systems are chosen to be 0.5 and 1.0~fs, respectively. Temperature is 
set to 330 K by utilizing random numbers as the initial velocities of atoms. 
The neighbor list with a 2~\AA\ buffer region 
is updated  every 50 MD steps. The thermodynamic data including the kinetic energy, 
potential energy, temperature, pressure are collected and recorded in every 50 MD steps.

\section{Machine configuration}  \label {sec:machine}

All numerical tests are performed on two supercomputers: Fugaku and Summit. 
The Fugaku supercomputer has 157,986 computing nodes and currently ranks 
No.1 in the top 500 list~\cite{Top500list} with a theoretical peak performance of 537 PFLOPS.
Each computing node is equipped with one A64FX, which has 4 internal groups 
named core memory groups (CMGs). Each CMG consists of 13 processor cores 
(one for OS activities), an L2 cache, and a memory controller. 
Each CMG has 8GB second-generation high-bandwidth memory (HBM2), so one 
A64FX has 32 GB HBM2 in total and the bandwidth is 1024GB/s. 
The A64FX supports the Scalable Vector Extension (SVE) and
the vector length of SVE can be 128, 256, and 512 bits. 
Theoretical peak performance of A64FX is 3.07/3.38 TFLOPS double-precision 
operation at 2.0 GHz and 2.2 GHz (auto boost), respectively.
The tests on Fugaku are performed with an MPI+OpenMP hybrid parallelization scheme, all threads within an MPI task are bound to an individual NUMA node to take advantage of the CPU-memory affinity. 

All GPU-related tests are performed on the Summit supercomputer, which
ranks No. 2 in the top 500 list~\cite{Top500list} with a theoretical
peak performance of 200 PFLOPS. 
Summit has 4,560 computing nodes, and each node has 2 identical groups 
consisting of one IBM POWER 9 socket and 3 NVIDIA V100 GPUs. The two
groups of hardware are interconnected via X-Bus with a bandwidth of
64 GB/s. Each IBM POWER 9 socket has 22 CPU cores and is equipped 
with 256 GB main memory with a bandwidth of 135 GB/s. 
Each V100 GPU has 16 GB of HBM with a bandwidth of 900 GB/s, 
and its double-precision theoretical peak performance is 7 TFLOPS.
The GPUs within a single group are connected through NVLink with 
a bandwidth of 50 GB/s. A non-blocking fat-tree topology using dual-rail Mellanox EDR InfiniBand with a bandwidth of 25 GB/s is formed among computing nodes, with one network adapter connected to one group of hardware. 
We use 6 MPI tasks per computing node in all tests on Summit by binding 
3 MPI tasks to each group of hardware (each MPI binds to an individual GPU)
to fully exploit the CPU-GPU affinity and network adapter. 

\section{PERFORMANCE RESULTS} \label{sec:performance}

In this section, we test the performance of the DeePMD-kit first on a single GPU and A64FX, and then the scaling behavior on Summit and Fugaku. Note that in all comparisons, the baseline code is the current state-of-the-art~\cite{2020Pushing} version of DeePMD-kit. 

\begin{figure}
    \centering
    \includegraphics[width=0.45\textwidth]{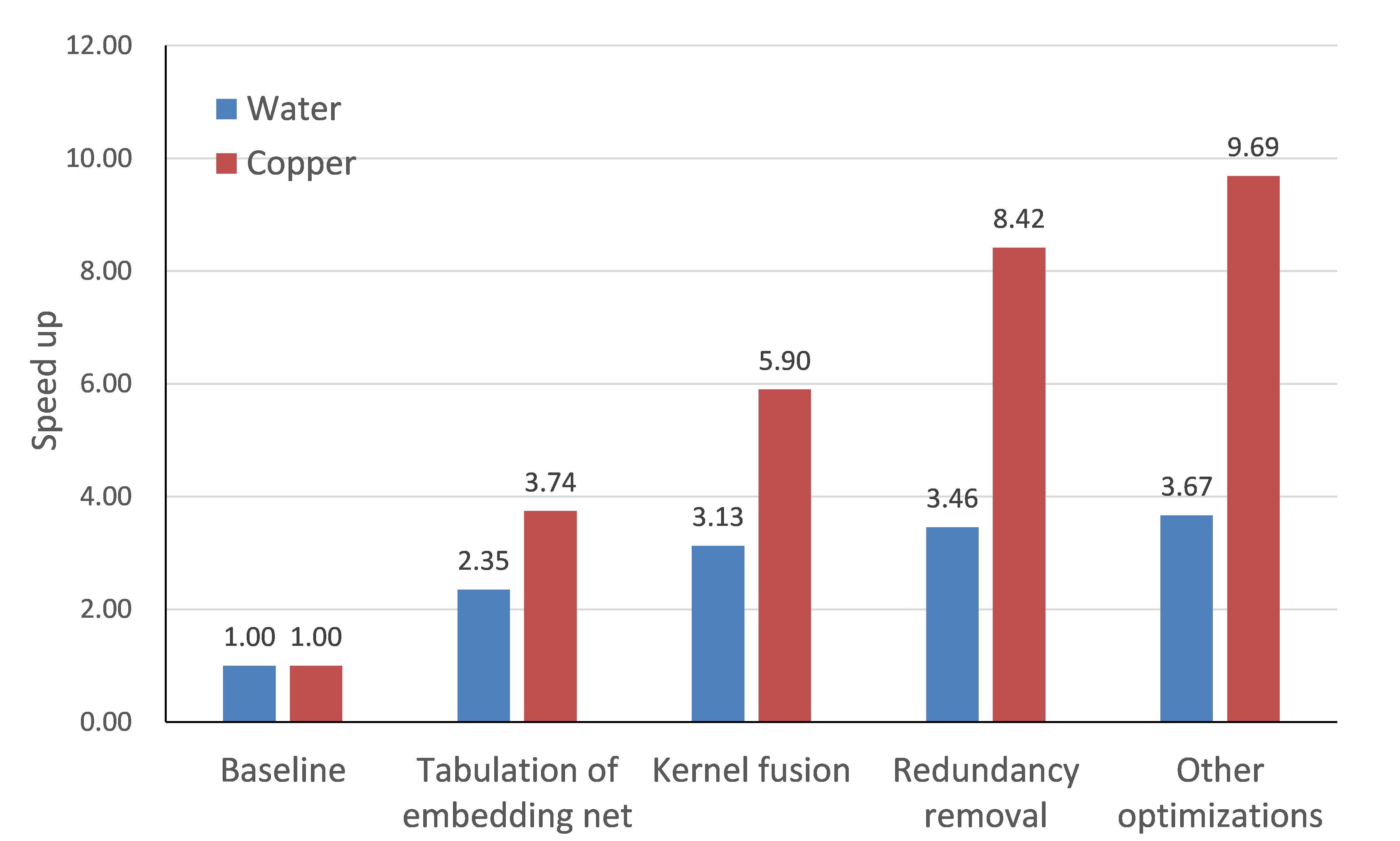}
    \caption{Step-by-step performance improvement on a single V100 GPU machine using algorithmic and system innovations. Baseline is the results from Ref.~\cite{2020Pushing}.}
    \label{fig:Single GPU}
\end{figure}

\subsection{Single V100 GPU}

The performance results on $12,880$-atom water and $6,912$-atom copper systems are tested to measure the performance improvement of critical algorithmic and system innovations on a single V100 GPU. The speedup of the time-to-solution for 99 steps of MD based on a step-by-step optimization is shown in Fig.~\ref{fig:Single GPU}.

\subsubsection{Tabulation of embedding net}\label{sec:algorithmic_result}

Compared to the baseline, we find that the optimized DeePMD-kit can be $2.3$ and $3.7$ times faster for water and copper system after introducing the tabulation of embedding net.
Note that {$82$ percent of FLOPS} (see Sec.~\ref{sec:algorithm_innovation}) are saved by the tabulation, implying a 5.6 times  speedup.
The actual speedup is less than ideal because DeePMD-kit is memory-bound rather than compute-bound~\cite{2020Pushing}, the speedup factor is determined by the memory-access reduction in the inference. 


\subsubsection{Kernel fusion}\label{sec:kernel_fusion_result}

As detailed in Sec.~\ref{sec:kernel_fusion}, both memory footprint and computational time are reduced by merging the tabulation and consecutive matrix-matrix multiplication into a single customized CUDA kernel. 
Testing results show that the maximum number of atoms accommodated on a single V100 GPU increase by a factor of {6 and 26} for water and copper systems, respectively. 
The total speedup is $3.1$ and $5.9$ compared to the baseline for water and copper, respectively. 

\subsubsection{Redundancy removal}\label{sec:redundancy_result}

As discussed in Sec.~\ref{sec:redundancy}, the evaluation of redundant elements is skipped to reduce the floating-point operations. 
Compared with the baseline, the speedup factor increase to $3.4$ and $8.4$ for water and copper systems, respectively. 
Profiling results show that our optimized kernel achieves $94\%$ of the 900 GB/s bandwidth provided by the V100 GPU.
We remark that copper gains a higher speedup in this step due to a higher degree of redundancy, as discussed in Sec.~\ref{sec:redundancy}. 



\subsubsection{Other optimizations}

The customized TensorFlow operators in DeePMD-kit are also optimized, and the corresponding speedup compared to the baseline is shown in Fig.~\ref{fig:Single GPU}. Overall, the time-to-solution of the optimized DeePMD-kit is 3.7 and 9.7 times faster, and system size increase by a factor of 6 and 26 on a single V100 GPU for water and copper, respectively.

\subsection{Single A64FX} \label{sec:a64fx}
    
\begin{figure}
    \centering
\includegraphics[width=0.5\textwidth]{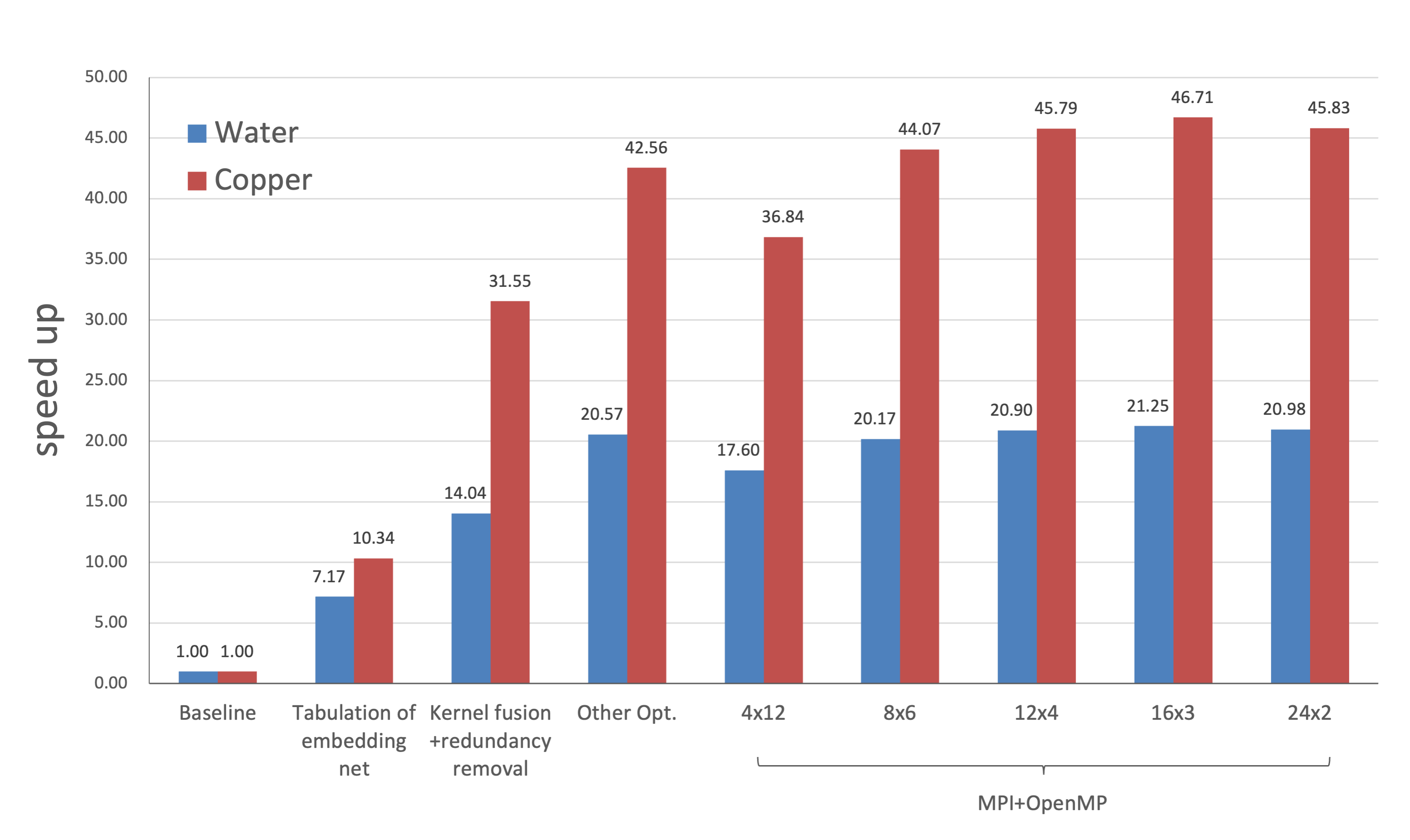}
    \caption{Step-by-step performance improvement on a single A64FX node for the $18,432$-atom water and $2,592$-atom copper systems.}
    \label{fig:Single A64FX speed up}
\end{figure}

Two systems, $18,432$-atom water and $2,592$-atom copper, are tested to measure the performance of the optimized DeePMD-kit on a single Fugaku node (A64FX). Note that the baseline is a flat MPI version of DeePMD-kit on A64FX. We remark that the GPU baseline is highly optimized ~\cite{2020Pushing}, but the A64FX baseline is not optimal and requires further optimization.  A step-by-step optimization is detailed at Sec.~\ref{sec:system_innovation2}, and the corresponding speedup is shown in Fig.~\ref{fig:Single A64FX speed up}.


\subsubsection{Tabulation of embedding net}

The data layout of the tabulated polynomial coefficients is rearranged to take
advantage of the 512-bit SVE instructions. 
As shown in Fig.~\ref{fig:Single A64FX speed up}, after rearranging the data layout of the tabulated polynomial coefficients discussed in Sec.~\ref{sec:tabulation_layout}, the optimized model can be $7.2$ and $10.3$ times faster compared to the baseline
for water and copper, respectively.


\subsubsection{Kernel fusion and redundancy removal}

We find that the kernel fusion and redundancy removal not only brings significant performance improvement on GPU but also work efficiently on CPU architecture. As shown in Fig.~\ref{fig:Single A64FX speed up}, the optimized model can be $14$ and $31.5$ times faster compared to the baseline for water and copper, respectively. 

\subsubsection{Other Optimizations}\label{sec:other_opt_results}

The customized TensorFlow operators, such as ProdEnvMatA op, ProdVirialSeA op and ProdForceSeA, are vectorized in the optimized DeePMD-kit, as detailed in Sec.~\ref{sec:vectorization}. We achieve an average speedup factor of $1.2$, $8.3$ and $7$ for ProdEnvMatA op, ProdVirialSeA op and ProdForceSeA, respectively. The corresponding proportion of the customized operators decreases from $15\%$ and $25\%$ to $10\%$ and $17\%$ for water and copper, respectively. 



Tanh is the activation function used in the DP model, and it takes $32\%$ and $20\%$ of total computational time for water and copper, respectively, on A64FX. In the optimized DeePMD-kit, Tanh is accelerated by tabulation as detailed in Sec.~\ref{sec:vectorization}. We achieve a speedup of $66.4$ and $44.5$ times for water and copper systems without any loss of accuracy. As shown in Fig.~\ref{fig:Single A64FX speed up}, the overall speedup factor reaches $20.5$ and $42.5$ for water and copper systems, respectively. 


\subsubsection{MPI+OpenMP}


MPI+OpenMP parallelization is introduced into the optimized DeePMD-kit to save memory usage and inter-MPI communication on the many-core architecture, as discussed in Sec.~\ref{sec:MPI+openMP}. The extra copies of the TensorFlow graph and system buffers are no longer needed.
The system size accommodated on single A64FX node increase from
$110,592$  to $165,888$  when using a 16$\times$3 configuration for water. Note that because the TensorFlow graph for the copper system is small (13 MB),  
the system size of copper using MPI+OpenMP almost remains the same as the flat-MPI version. Testing results show that  MPI+OpenMP implementation can also be slightly faster than the flat MPI except when using 4$\times$12 (Each MPI task consists of 12 threads, and resides on an individual CMG), as shown in Fig.~\ref{fig:Single A64FX speed up}. Since the 16$\times$3 configuration is the optimal choice both in system size and computational speed, we use 16$\times$3 as the configuration in the scaling tests.

\subsection{Comparison between A64FX and V100}

\begin{table}[t]
\caption{Comparison between A64FX and V100: the time-to-solution (TtS) 
(in us/step/atom), the normalized time-to-solution with respect to FLOPS (TtS$\times$Peak), and the normalized TtS with respect to power consumption (TtS$\times$Power) are listed. Numbers in brackets are speedup factors as compared to the V100 results. 
}
\begin{tabular}{llccc}
\toprule
                        &        & 
                        {TtS}  &  
                        {TtS $\times$ Peak}   &  
                        {TtS $\times$ Power}  \\ 
\midrule
\multirow{2}{*}{{Summit}} & 
    {water}  & {2.58} & {18.1(1.0)}  & {952.0(1.0)}  \\
            &   
    {copper} & {2.87} & {20.1(1.0)}  & {1059.0(1.0)} \\ 
\midrule

\multirow{2}{*}{{Fugaku}} & 
    {water}  & {4.47} & {15.1(1.2)} & {737.6(1.3)} \\
            & 
    {copper} & {5.78} & {19.5(1.03)} & {953.7(1.1)} \\ 
\bottomrule
\end{tabular}

\label{tab:comparison between A64FX and V100}%
\end{table}

For the water system, we achieve 4.47 us/step/atom with one A64FX and 2.58 us/step/atom with one V100 GPU. The time-to-solution is normalized for a fair comparison, as shown in Table~\ref{tab:comparison between A64FX and V100}. First, we normalize the time-to-solution with respect to peak performance by multiplying time-to-solution with the theoretical peak of V100 and A64FX. We find that the normalized results on A64FX can be $1.2$ and $1.03$ times faster than V100 for water and copper systems, respectively. The average power consumption of a single V100 and A64FX is 369 and 165 watts~\cite{Top500list}, respectively. Then the time-to-solution is normalized with respect of power consumption, as
shown in Table~\ref{tab:comparison between A64FX and V100}. By setting V100 as the baseline, the speedup factor of A64FX is 1.3 and 1.1 for water and copper, respectively. 

\subsection{Scaling}

The scalability of our optimized DeepMD-kit is evaluated on Summit and Fugaku. Note that a fraction of the Fugaku supercomputer is used due to the computational resource accessible to us. The system size can reach an unprecedented {$17$} billion atoms, {$134$} times bigger compared to the current state-of-the-art. And the corresponding time-to-solution can be {$3.7$ to $9.7$} times faster compared to Ref.~\cite{2020Pushing}. Note that in all scaling tests, we use an MPI+OpenMP parallelization configuration of $16\times3$. 


\subsubsection{Strong Scaling.} \label{sec:strong_scaling}



\begin{figure}[hb]
    \centering
    \includegraphics[width=0.45\textwidth]{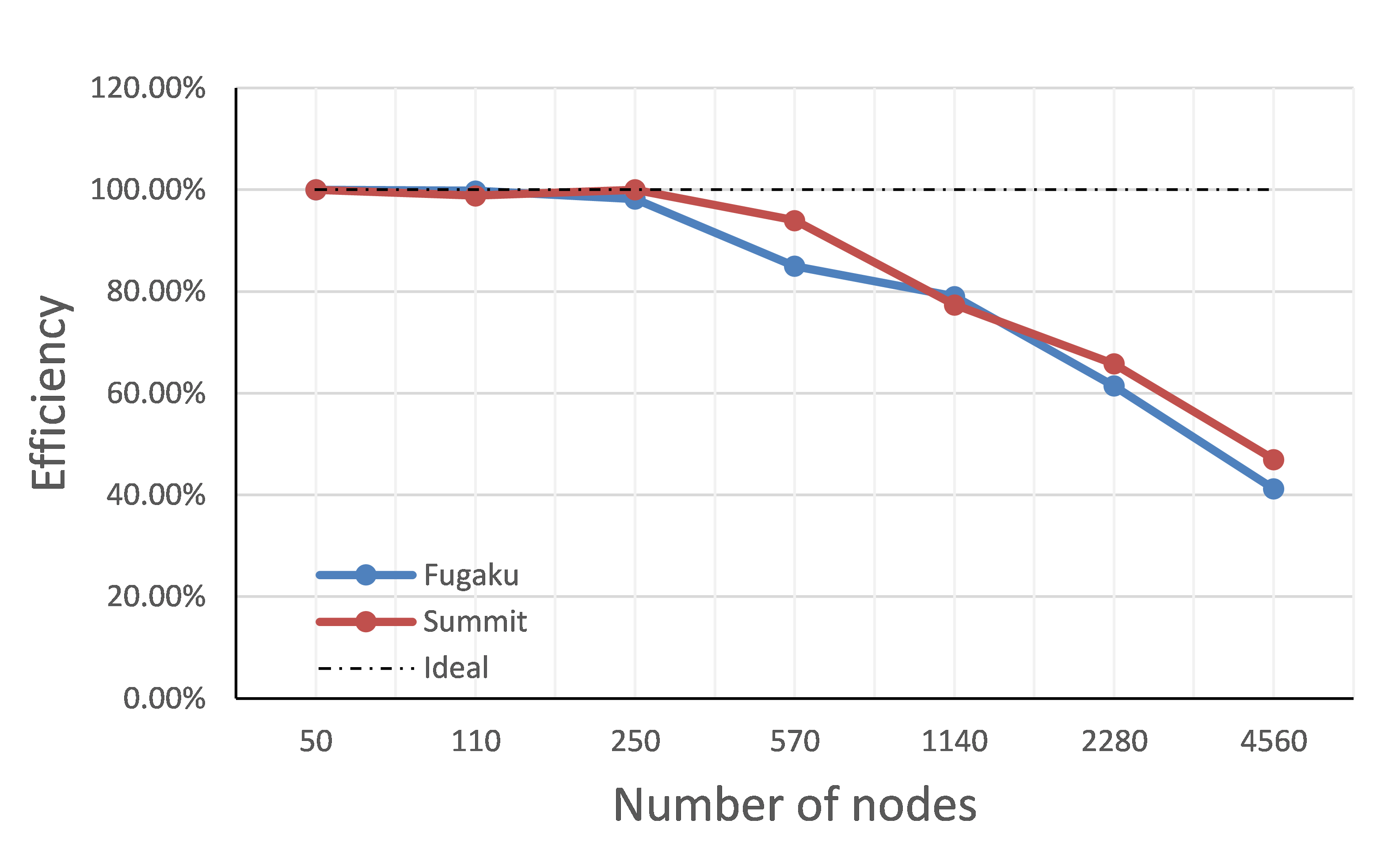}
    \caption{Strong scaling of MD simulations (99 steps, atomic forces and the total energy are computed 100 times) for the water system: a total of $8,294,400$ atoms on Fugaku and $41,472,000$ atoms on Summit are used.}
    \label{fig:strong scale water}
\end{figure}
\begin{figure}[hb]
    \centering
    \includegraphics[width=0.45\textwidth]{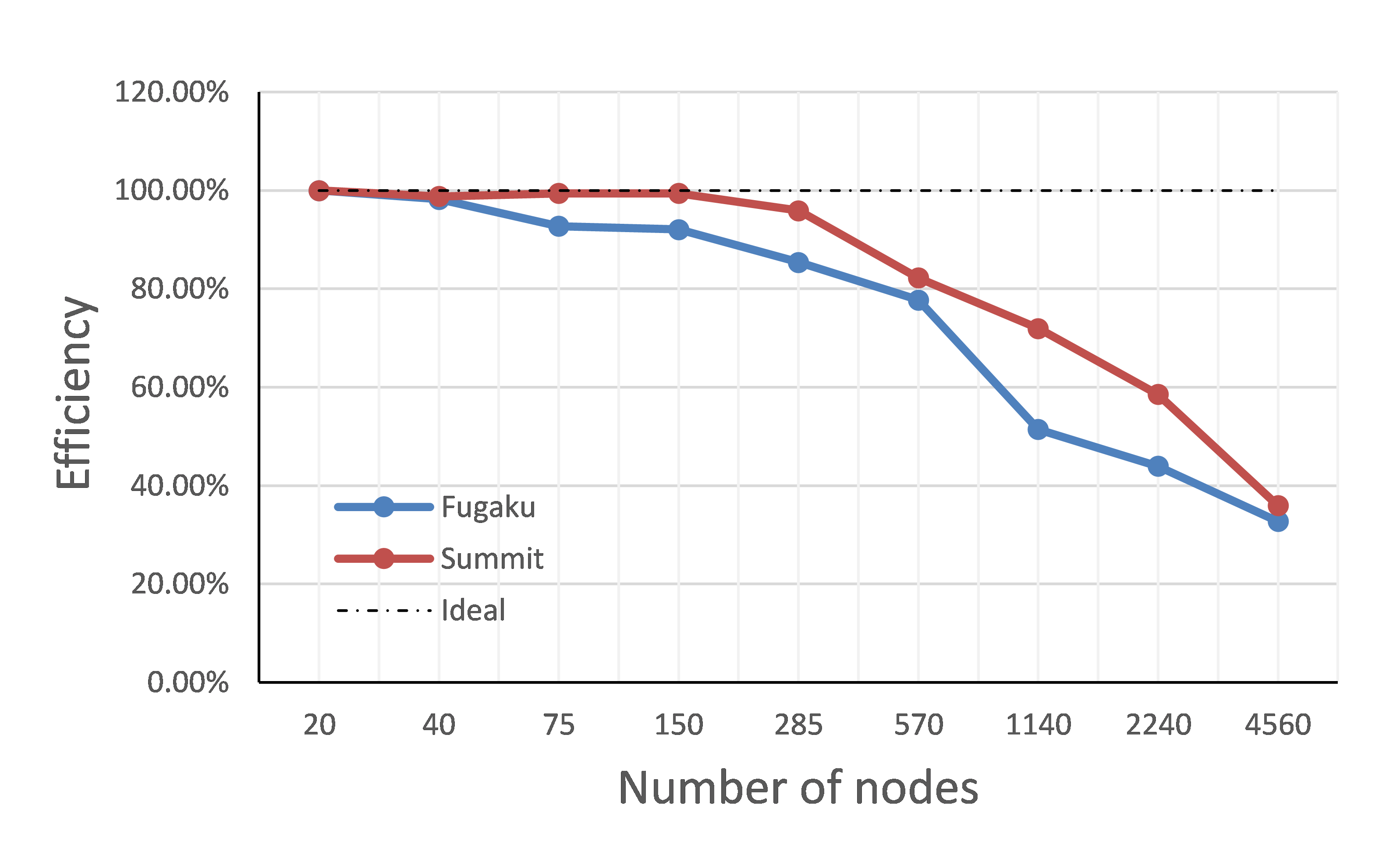}
    \caption{Strong scaling of MD simulations (99 MD steps, atomic forces and the total energy are evaluated 100 times) for the copper system: a total of $2,177,280$ atoms on Fugaku and $13,500,000$ atoms on Summit are adopted in these MD simulations.}
    \label{fig:strong scale copper}
\end{figure}


We measure the scalability of the optimized DeePMD-kit using the parallel efficiency of water ({$8,294,400$} atoms for Fugaku, $41,472,000$ atoms for Summit) and copper ({$2,177,280$} atoms for Fugaku, $13,500,000$ for Summit) systems. 
The scaling behavior ranging from $20$ to $4,560$ computing nodes of Summit and Fugaku are nearly the same, as shown in Fig.~\ref{fig:strong scale water} and Fig.~\ref{fig:strong scale copper}. Both water and copper show nearly perfect scaling on up to $570$ and $285$ computing 
nodes, and can further scale to $4,560$ computing nodes. For the water system,  the parallel efficiency is $46.99\%$ on $4,560$ computing nodes on Summit, and that of the Fugaku is $41.20\%$. The corresponding time-to-solution is {$6.0$} and {$2.1$} nanoseconds per day
on Summit and Fugaku, respectively. 
For the copper system, the parallel efficiency on $4,560$ Summit nodes 
is  $35.96\%$, and the corresponding parallel efficiency is $32.76\%$ for Fugaku. The corresponding time-to-solution can reach {$11.2$} and {$4.7$} nanoseconds per day. 

We notice that the parallel efficiency on Summit is slightly better 
compared to that on Fugaku due to its higher ratio in computation over communication. 
Since the computational complexity of DP method is linear ($O(N)$), 
the ratio of computation over communication can be approximated with the 
size of the sub-region over the size of the ghost region. 
Note that we launch 16 MPI tasks on single A64FX node, but only 6 MPI process
on one Summit node. When scaling to $4,560$ computing nodes, $72,960$ MPI tasks and 
$27,360$ MPI tasks are launched on Fugaku and Summit, respectively. 
In the strong scaling of the copper system, each MPI task on Fugaku holds a 
sub-region of $113$ atoms, and its ghost region is $1,735$. Meanwhile the corresponding 
number on Summit is $1,515$ and $7,520$, respectively. The ratio of the computation 
over communication is $\nicefrac{1}{15}$ and $\nicefrac{1}{5}$ on Fugaku and Summit, respectively. 

\subsubsection{Weak Scaling.} \label{sec:weak_scaling}

\begin{figure}
    \centering
    \includegraphics[width=0.45\textwidth]{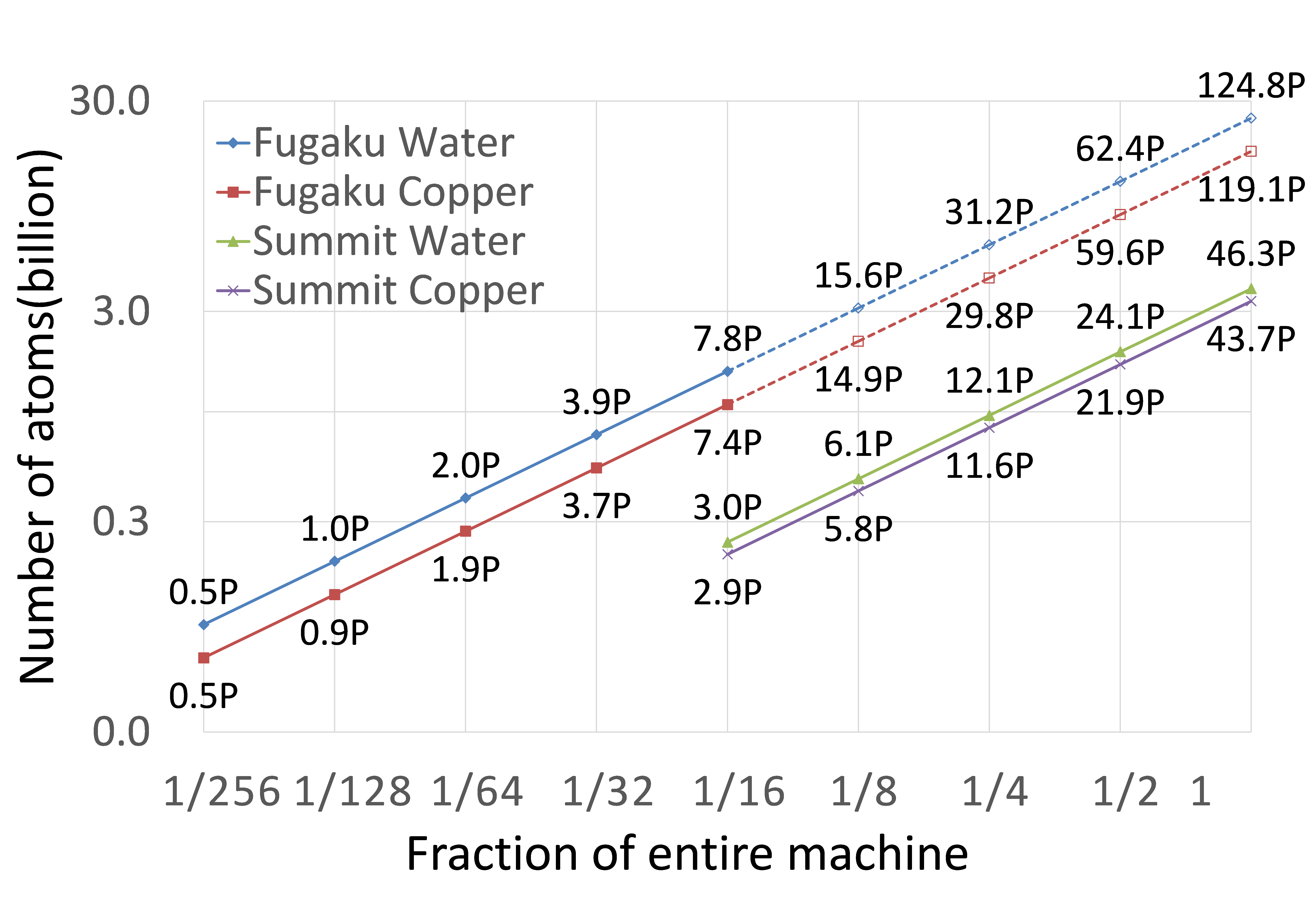}
    \caption{Weak scaling of water and copper systems ranging from $\nicefrac{1}{256}$ to 
    the entire machine (Summit and Fugaku). Note that the dotted line 
    for Fugaku is a projected result due to the limited computing resources.}
    \label{fig:weak scaling}
\end{figure}

The weak scaling of the optimized DeePMD-kit is measured in terms of the system size and FLOPS of 99 MD steps for both water and copper systems (Fig.~\ref{fig:weak scaling}).
As shown in Fig.~\ref{fig:weak scaling}, both systems show perfect scaling with respect to the number of nodes used on Fugaku and Summit. On Fugaku, the maximal number of atoms for water and copper can reach $1.56$ and $1.08$ billion atoms, respectively, on $9,936$ nodes, and a projected estimation shows that the DeePMD-kit can reach $24.9$ and $17.3$ billion atoms for water and copper, respectively. 
We remark that our estimation of the weak scaling is reasonable based on the communication 
pattern of the DeePMD-kit. 
For the copper system, the time-to-solution  can reach $4.1 \times 10^{-11}$ seconds/step/atom, and the corresponding peak performance achieved is $119$ PFLOPS ($22.17\%$ of the theoretical peak). Compared to the current state-of-the-art, the projected system size is 134 times bigger and the corresponding time-to-solution can be $20$ times faster. 
On Summit, the maximum number of atoms can reach  $3.9$ and $3.4$ billion atoms for water and copper, respectively. 
For the copper system, the time-to-solution can reach $1.1 \times 10^{-10}$ seconds/step/atom, and the corresponding peak performance achieved is $43.7$ PFLOPS ($22.8\%$ of the theoretical peak). 
Compared to the current state-of-the-art, the projected system size is $27$ times bigger and the corresponding time-to-solution can be $7$ times faster. 
Due to the perfect linear scaling, we foresee that our optimized DeePMD-kit code can compute larger physical systems on near-term and future exascale supercomputers without essential difficulties.





\section{Conclusion and Future work} \label{sec:conclusion}


In this paper, we presented an optimized version of DeePMD-kit by adapting a novel tabulated DP model and system optimizations on the top two supercomputers: Summit and Fugaku. The tabulated model reduces the floating-point operations in the MLMD, and the consecutive optimizations improve the data locality/movement and memory usage on both CPU and GPU. Compared to the current state-of-the-art, our optimized code extends the capacity of MLMD to 10 billion atoms, with a time-to-solution of $1.1 \times 10^{-10}$ seconds/step/atom (7 times faster).
This work opens the door for unprecedentedly large-scale molecular dynamics simulations based on {\it ab initio} accuracy, and can be potentially utilized in studying more realistic applications such as mechanical properties of metals, semiconductor devices, batteries, etc.
The performance of the optimized version of DeePMD-kit is demonstrated on both Summit and Fugaku, and our optimization strategies can also be beneficial for other architectures, such as Intel CPUs and the AMD GPU supercomputer Frontier targeting at exascale computing. The combination of the compressed neural network model and cross-kernel dataflow optimizations provide insight in exploiting the computing power provided by the modern supercomputer, especially for HPC+AI applications.  The mixed-precision versions of code still has accuracy problems and will be our future work. 

\begin{acks}
The numerical experiments are performed on Fugaku and Summit supercomputers. The computational resources of supercomputer Fugaku are provided by the RIKEN Center through the HPCI System Research project (Project ID: hp200307). The computational resources of supercomputer Summit are provided by the U.S. Department of Energy through its Innovative and Novel Computational Impact on Theory and Experiment (INCITE) program (Project ID: CHP115). This work is supported by the following funding: National Science Foundation of China under Grant No. (61802369, 61972377, 62032023, 12074007, 12122401), CAS Project for Young Scientists in Basic Research(YSBR-005), GHFund A(No. 20210701), Director's Funding of Key State Laboratory of Computer Architecture(E04118), Network Information Project of Chinese Academy of Sciences(CAS-WX2021SF-0103), and Huawei Technologies Co., Ltd..
\end{acks}

\bibliographystyle{ACM-Reference-Format}
\bibliography{main}

\appendix









\end{document}